# UV ASTRONOMY WITH SMALL SATELLITES

**Pol Ribes-Pleguezuelo[(1)], Fanny Keller[(1)], Matteo Taccola[(1)]**

[(1)] *ESA-ESTEC, Keplerlaan 1, 2201AZ Noordwijk, Netherlands,*
*pol.ribes@esa.int, fanny.keller@esa.int, matteo.taccola@esa.int*

## ABSTRACT

Small satellite platforms with high performance avionics are becoming more affordable. So far, with a few exceptions, small satellites have been mainly dedicated to earth observation. However, astronomy is a fascinating field with a history of large missions and a future of promising large mission candidates. This prompts many questions; can the recent affordability of small satellites change the landscape of space astronomy? What are the potential applications and scientific topics of interest, where small satellites could be instrumental for astronomy? What are the requirements and objectives that need to be fulfilled to successfully address the astronomical investigations of interest? Which kind of instrumentation suits the small platforms and the scientific use cases best? This paper discusses possible scientific use cases that can be achievable with a relatively small telescope aperture of 36 cm, as an example. The result of this survey points to a specific niche market -astronomy observation in the UV spectral range. UV astronomy is a research field which has had valuable scientific impact. It is, however, not the focus of many current or past astronomical investigations. UV astronomy measurements cannot be made from earth, due to atmospheric absorption in this spectral range. Only a few current space missions, such as the Hubble and Gaia, cover the UV spectral range, some of them only in the near-UV (NUV). The research field is currently sparsely addressed but of scientific interest for a large community. Small satellites offer the opportunity to provide more means of research for UV astronomy. Therefore, this paper also presents an instrument design with a modest telescope aperture, a spectrometer and a detector that is suitable for observations in the UV. The observatory design can be accommodated on small platforms for the selected scientific use cases. It fulfills the scientific objectives and requirements of those use cases. The paper also discusses how effectively the instrument design allows to investigate the selected scientific applications.

## 1   INTRODUCTION

In 2021, after many years of preparation, the multibillion dollar James Webb Space telescope (JWST) will finally be launched to start its mission of studying early formed stars and galaxies. The telescope, designed and built thanks to the partnership of many countries, was designed for infrared analysis. Thought also as a replacement for the Hubble Space telescope (HST), it however, has a lack of UV wavelength detection capabilities; an interesting regime which recently, seems to be less considered when new space telescope missions are proposed.

UV space astronomy can provide relevant information to study the chemical and thermal compositions of bodies in space, while this information must be collected outside of earth's atmosphere due to the high atmospheric absorption in this wavelength range. Previous UV telescopes, such as the HST, the UV telescopes used for the Space Shuttle Astro missions (launched in the 80s and 90s decades) [1] or the Galaxy Evolution Explorer or GALEX (launched in 2003), provided useful data for scientists to understand space plasma dynamics, or to separate interstellar spectral lines [2]. However, most of these instruments are nowadays no longer accessible or are about to be decommissioned.

UV astronomy can be used to understand aurora and atmospheric dynamics in planetary science, for the study of absorption lines of interstellar and intergalactic media, to perform the spectroscopy of stars to study their winds and evolution or for cosmology purposes, amongst others [2]. The lack of UV instruments being developed for future missions, new technological advances in instruments and materials enabling the designing of improved telescopes and the reducing cost of launching CubeSats [3], has prompted this feasibility study for the design and construction of modest-cost UV telescopes which could help in the further understanding of the universe.

## 1.1 Background

At the end of the 20th century, many UV space telescopes had been planned and launched but today, some of them have been decommissioned without replacement. A few examples of the last functional UV space telescopes are:

The **HST UV** that contains the so-called Space Telescope Imaging Spectrograph (STIS) instrument; repaired in 2009 after a 2004 failure. It has near and far-UV capabilities between 115-310 nm with a resolving power of 500-1000 [4]. Later, during the 2009 instrument reparation, a new instrument called the Cosmic Origins Spectrograph (COS) was installed. The COS design allows for 90-320 nm studies with a resolving power between 1500-24000 [5]. The HSP is planned to be operative until the mid-2020s [6].

The **Space Shuttle UV telescopes.** Several UV telescopes were planned to be launched with the Space Shuttle during the 80s and 90s decades. However, due to the Challenger accident and the following Space Shuttle budget funding cuts, most of the UV missions were not continued and only four of the planned UV telescopes made it into space. One of the telescopes, which could fly, was the Ultraviolet Imaging Telescope (UIT). The device had a 38 cm Cassegrain design that operated in the far (120-200 nm) and mid-UV (200-320 nm) [7].

The **Galaxy Evolution Explorer (GALEX)**, launched in 2003 with the goal of studying star formation processes with a working wavelength range of 135 to 280 nm [8]. In this case, the telescope was of a Richey- Chrétien design with a 500 mm aperture [9], with a resolving power of 100 in the NUV [10].

Additionally, a couple of large UV telescope missions are in the process of being designed and implemented, those are:

The **World Space Observatory (WSO)**, which has required a long period of time to be approved - the first concepts were drafted in 1999. The mission is run by a Russian/Spanish consortium and is finally going to be launched in 2023. The instrument will study wavelengths between 115 and 315 nm with a resolution power between 1000 to 50000, using a Ritchey–Chrétien telescope [11].

The **Cosmic Evolution Through Ultraviolet Spectroscopy (CETUS)**, originally a NASA Probe Class mission, with a wide-field, 1.5 m aperture telescope. The mission will cover the UV from 100 to 400 nm by using three different scientific instruments (near-UV multi-object slit spectrograph (MOS), a wide-field near-UV and far-UV camera, and a near-UV and far-UV single object Spectrograph [12]).

Due to the void in UV coverage over the next decade, some UV missions are being expedited, for example, the WSO and CETUS. These missions are however, relatively expensive, would require some years until they are available for the scientific community to use, and they can only cover specific wavelengths, resolutions and mission purposes. In order to be able to have broadly accessible UV experiments outside Earth in a short period of time, we propose the possibility to design, manufacture and launch small UV telescope satellites with a resolution power of between 1000-10000, in order to study exoplanets (as an example, as this is currently an astronomical topic of great interest) [11]. Similar projects should be able to be drafted, built and launched in a period of not more than three to four years with the current available technology and market-accessible products. Moreover, thanks to the new technology advances in optics manufacturing, missions are nowadays able to cover the far-UV (FUV) regime. The FUV is particularly interesting because it can be implemented to study exoplanet formation or exoplanet composition (as in Lyα), but also baryonic matter or large-scale structures [13]-[14]. These kind of missions should allow access to the scientific community to ensure that UV, but also FUV [15] science is still carried out.

## 2 PROPOSED DESIGN

### 2.1 Optical design

The initial proposed designs are based on a 36 cm telescope, so that the optical payload can fit within a 27U CubeSat platform. Moreover, in order to investigate exoplanets in the UV regime in transit processes, the telescope should be able to acquire $10^{-3}$ to $10^{3}$ W/m$^2$/μm [16] star irradiance and distinguish the 1-10% decreased signal when an exoplanet-star transit occurs [17]. If the project has a goal of studying exoplanet atmospheric composition using a 120 to 280 nm wavelength signal, the spectral resolution should also be between 0.5 to 1 nm.

#### 2.1.1 Telescope

What we propose for the telescope design is to use a 2-mirror Cassegrain with a rectangular aperture which can be built using additive manufacturing techniques. The primary mirror would have a radius of 570 mm and a 360 x150 mm rectangular size, while the secondary mirror would have a radius of 246 mm and a 120 x 48 mm rectangular size. The design would fit in a 27U CubeSat structure, by minimizing the amount of components, and by making the whole system as light and as compact as possible. In Figure 1, the mirror sketch in Zemax shows a rectangular aperture telescope that could be an initial design baseline. The design uses just two mirrors in order to decrease instrument mass.

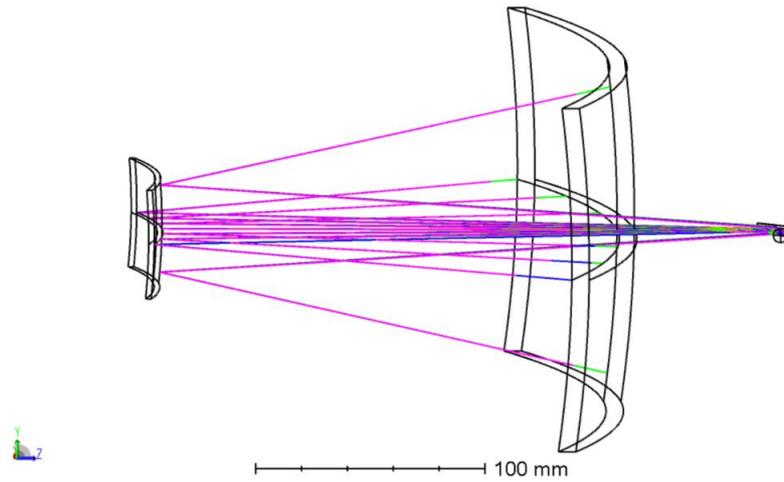

*Figure 1 2-mirror simplified design telescope. The secondary mirror is on the left and the primary mirror is on the right. The design uses conic mirrors with an approximate telescope aperture of 36 cm.*

### 2.1.2 Instrument

The instrument will include a UV spectrograph capable of operating between 120 and 280 nm with a resolving power between 1000 and 10000. A modest-cost and low weight UV instrument should have a minimum amount of optical components. A way to reduce the amount of optical components is to use a spectrograph with the grating embedded in a spherical concave surface which provides a single-reflection surface-instrument. This is beneficial when working in the FUV, since optical components have lower reflectivity in the FUV. A similar approach is also being used for the COS instrument in the HST [18]. In this last case, the holographic ruled gratings were provided by the company Horiba. Figure 2 shows the Zemax sketch of the spectrograph that we propose for the baseline design. A folding mirror brings the light from the telescope secondary mirror into the spherical grating, which then diffracts the light onto the detector. Today, commercial UV holographic gratings can be embedded in spherical surfaces to obtain 2000 to 5000 lines/mm, which could satisfy the mission demands [19],[20].

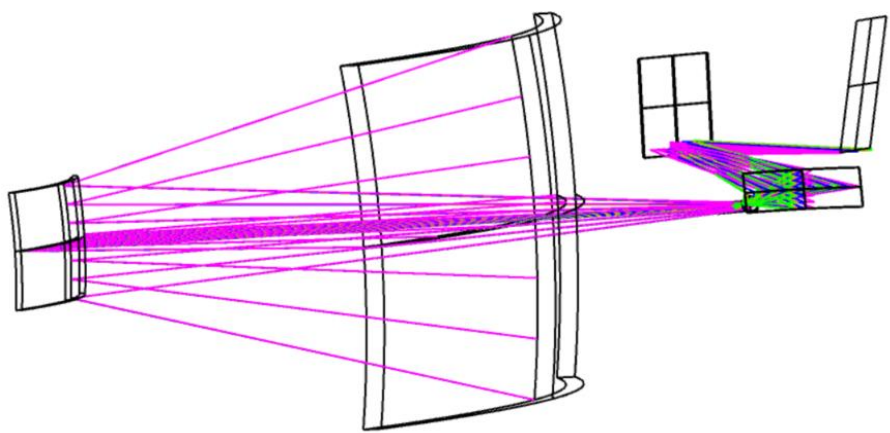

*Figure 2 On the left side of the figure, the designed telescope as seen in Figure 1 (with primary and secondary mirrors) is shown; and on the right, the spectrometer and detector are shown.*

It should be noted that the proposed baseline design is based on incoming star irradiance requirements but not image quality, which is not a driving parameter for studying exoplanets in the transit regime. An example of a similar approach is the CHaracterising ExOPlanet Satellite (CHEOPS) ESA mission, where the analysis is performed using a largely defocused image of the stars [21].

### 2.1.3 Detector

The efficient detection of FUV could be a difficulty for the proposed research, since conventional detectors such as back illuminated CCDs have a poor response at these wavelengths (around 20%). However, micro-channel plate (MCP) detectors could be an alternative for extreme UV regimes, as they have also been used in the HST COS instrument and in the Far Ultraviolet Spectroscopic Explorer (FUSE) mission. For the closer NUV (> 250 nm) wavelengths, back illuminated CCD detectors could be used.

### 2.1.4 FUV Coatings

An important aspect to consider when working in the UV regime, and more specifically in FUV astronomy, is the lack of proper UV coatings that can be applied on mirror surfaces giving high reflectivity and a wide wavelength bandpass. Examples of this problem could be observed in FUV missions launched in the last decades where the applied layers reflectivity was in some cases only 40%. Although our design tackles this issue by reducing the amount of reflecting surfaces, in fact, high quality and highly reflective FUV coatings should still be a focus of research. In the last years, several groups have worked on UV coatings to improve the future of UV astronomy. NASA reports the use of aluminium mirrors based on protective $MgF_2$, $AlF_3$ and LiF coatings in [22],[23], achieving a reflectance up to 80% between 95 and 200 nm, by using advanced Atomic Layer Deposition (ALD) techniques [24]. Additionally, a different NASA group investigated mirrors made using Physical Vapour Deposition (PVD) processes, but by using similar protective layers and also obtained reflectances of over 80% [25]. NASA's goal is to be able to use these last mentioned coatings and techniques for the Large UV/Optical/IR Surveyor (LUVOIR) and CETUS space telescope missions. Similar initiatives are also being carried out by European institutions such as CNES (the National Centre for Space Studies- France) [26], CSIC (the Spanish National Research Council-Spain) [27], or CNR (the National Research Council- Italy) [28], showing similar results as those presented by NASA, but using different methodologies. The enhanced FUV coatings of the European project are planned to be used in the future WSO mission. These results in new FUV coatings assure the feasibility of near future FUV astronomical space missions.

## 2.2 State-of-the-art manufacturing processes

One of the main differences between the previously constructed UV telescopes and the current possibilities is that the state-of-art manufacturing processes allow us to now build lighter mirrors (reducing their mass by up to 64% [29]) which are more compact, but can still have affordable precision for UV science requirements. The proposed guidelines to reduce mass and volume can be achieved via additive manufacturing process using, for example, AlSi40 or Scalmalloy base-materials. Moreover, the internal telescope mirror's main body, which is one of the heaviest components could be also designed with carved cells (Figure 3) to reduce its mass [29]. These mirrors, as reported in [30], could be later covered with Nickel-Phosphorus (NiP) layers, then fine polished by diamond turning and/or Magnetorheological Finishing (MRF). By following these last mentioned processes, the surface quality can have a low surface roughness between 0.4 to 0.8 nm RMS (Root Mean Square) and form deviation of between 70 to 80 nm PV (Peak-to-valley) [30]; which could provide the necessary surface quality for UV astronomy.

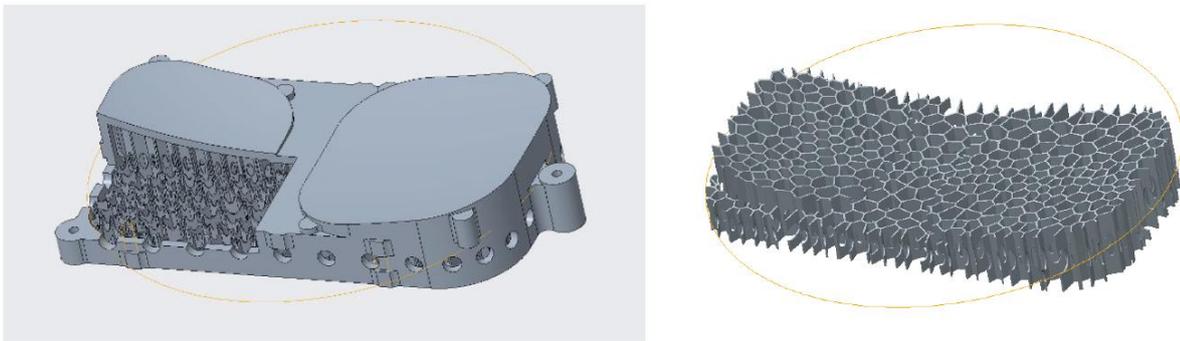

*Figure 3. Example of a lightweight manufactured mirror using additive manufacturing carved cells to build a freeform optical component. CAD model of a mirror showing a cut view (left). Carved cell lightweight internal structure (right). Source: Fraunhofer IOF [29].*

Moreover, articles from the same authors reported that similar techniques to minimize the volume and mass could be used for telescope housings or other instruments [31]-[32]. Such techniques have already been used in the manufacture of the Deutsches Zentrum für Luft- und Raumfahrt (DLR) Earth Sensing Imaging Spectrometer (DESIS) on board the International Space Station (ISS) [33]. The DESIS optical payload mass was around 21 kg [33].

## 2.3 Satellite Budgets

A current possibility for covering the lack of UV in a short period of time is the use of commercial off-the-shelf (COTS) satellite components. In this current study, we propose a 27U CubeSat with an approximate mass of 54 Kg and volume of 34 cm x 35 cm x 36 cm. However, the mass and volume constraints will have to take not just the optical payload (telescope, instrument and detector), but also other relevant parts of the satellite into account, such as:

- Satellite structure (harnesses, fasteners, mounting plates, etc.),
- Power supply (solar panels, batteries) and electronics,
- Position measurement (gyroscopes, start trackers, etc.) and position adjustment (reaction wheels, thruster etc.),
- Communication systems.

All those components are market accessible nowadays, space qualified and with space heritage and available for a relatively low cost. Examples of optical payload COTS for small satellites can be found in [34]. Similarly sized components required for the mission's success (including the designed telescope and instrument) can be seen sketched inside a 27U CubeSat platform in Figure 4.

The satellite will also have to function with a limited amount of power (~60 W) supplied and stored by the available solar panels and batteries. With these considered, the optical payload would have to be under an approximate mass of 25 Kg and work with a power consumption of about 45 W. Considering previously sent missions and used manufacturing technologies this budget seems to be in agreement with the proposed guidelines.

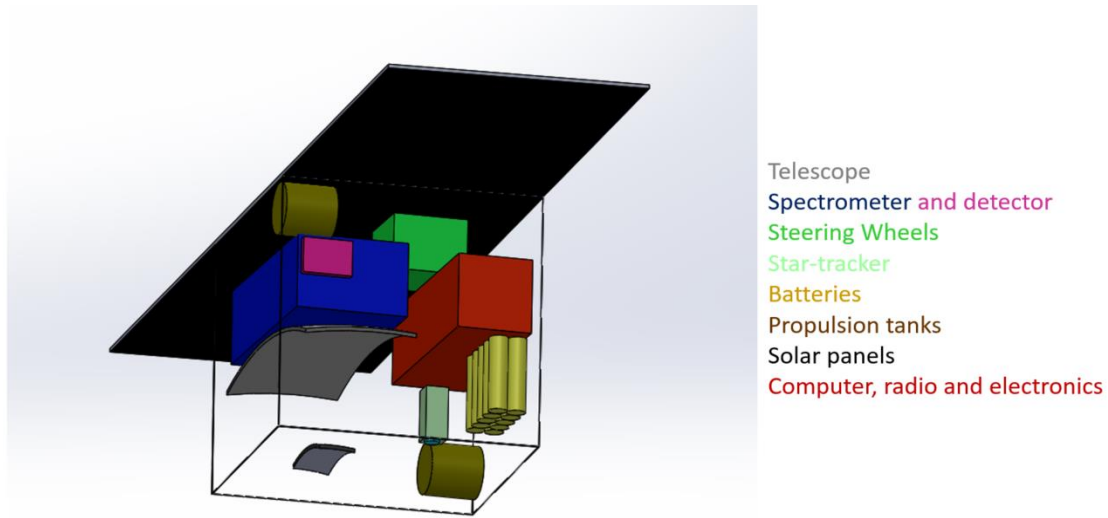

*Figure 4 Schematic of the optical payload and other mission-required devices mounted and fitted inside a 27U CubeSat platform.*

## 3 THE CUTE MISSION

During the present analysis the authors found a current UV small satellite mission in development by the University of Colorado; the Colorado Ultraviolet Transit Experiment or CUTE. The mission is a 4-year project funded by NASA and is planned to be launched in a 6U CubeSat platform in the second quarter of 2020. The project will use a Cassegrain telescope design covering the NUV (250-330 nm) with an instrument a resolving power of around 3000. The challenging experiment will include a 206 mm x 84 mm primary telescope mirror, a back-illuminated UV CCD detector and the total mission mass is said to be around 10 Kg. The scientific purpose of the mission is to study exoplanet atmospheric mass loss and magnetic fields [35].

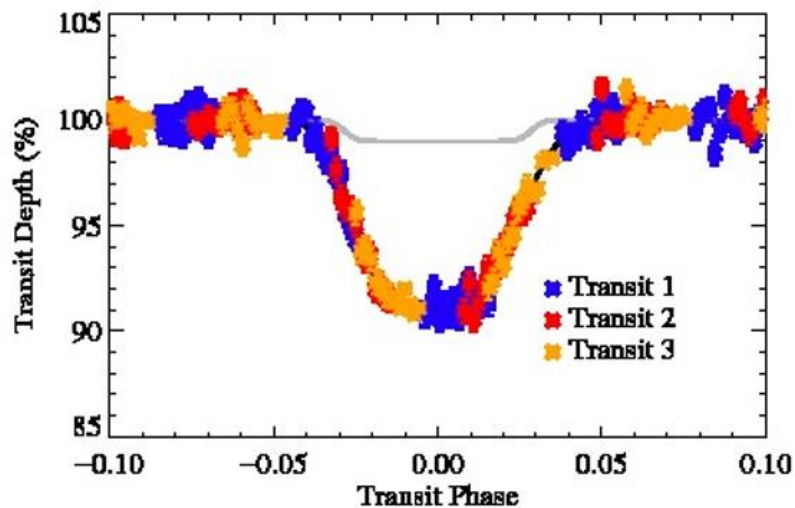

*Figure 5 CUTE simulated transit received light curve expected for targeted exoplanets. Source [36].*

## 4   DESIGN COMPARISON

The proposed baseline design shown in this feasibility study indicates that small and simplified telescope missions could be implemented in a short period of time to target a specific scientific need not currently covered by instruments in space. The proposed design was based on several previously launched missions such as the GALEX, a small UV telescope mission with an approximate mass of 280 kg; the UIT, having a 38 cm aperture telescope and covering similar wavelength ranges and on the FUSE and COS, for their FUV capabilities and spectrometer design. However, due to past technological constraints, all these previous missions required between one or two decades from design to use, had much heavier instruments and in some cases had much lower optical performances than that which can be achieved today using COTS products. Our starting baseline design (using newly developed technologies) allows a similar, or in some cases better, FUV analysis, with a much lighter and more compact design than the previously launched missions.

## 5   CONCLUSIONS AND OUTLOOK

Until now, most astronomical space telescopes have been proposed, designed, built and launched requiring a few decades until they could be operational. However, at the same time, new technology developments achieved a maturity that made some of the processes or even devices to be not the state-of-the-art in their field when the instruments were finally launched. Nowadays, most of the required parts of the instruments (solar panels and thrusters but also subsystems required for the optical payloads) are market available and can be provided in a period of a few months. In the current article, we showed how a small FUV mission could be designed with a compact size without losing performance when compared to past decommissioned missions.

During the research phase of this feasibility study, it was found a similar initiative had already been started, i.e. the CUTE mission. In this case, they designed a lightweight telescope that could even be fitted into a 6U CubeSat platform; showing that small telescope missions have already started paving their way into the future of space science.

Availability of small and easy to implement missions will allow the scientific community to be constantly working in different wavelength ranges and science fields. Thus, we see no reason why small CubeSat platforms with optical payloads need to be implemented only for Earth Observation missions [37], and why they cannot be used nowadays to study other bodies in the solar system or even beyond.

## 6   ACKNOWLEDGEMENTS

The authors want to give special thanks to ESA-ESTEC colleagues for their supervision and advice, in particular, Dr. Luca Maresi, Guido De Marchi and Dr. Alessandro Zuccaro Marchi. Also great thanks is due to colleagues from the Fraunhofer IOF, Dr. Frank Burmeister and Dr. Nils Heidler, for their advice in new manufacturing techniques for telescope mirrors and gratings. Finally, thanks to Dr. Rossá Gerard Mac Ciarnáin from Physicsproof.com for his proofreading service.